\documentclass[runningheads]{llncs}
\usepackage{graphicx}
\usepackage{times}
\usepackage{latexsym}
\usepackage{todonotes}
\usepackage{listings}
\usepackage{subfig}
\usepackage{multirow}
\usepackage{tabularx}
\usepackage{comment}
\usepackage{url}
\usepackage{subfig}
\usepackage{amssymb}
\usepackage{makecell}
\usepackage{enumitem}
\usepackage[layout=footnote]{fixme}
\graphicspath{ {./img/} }

\newcolumntype{Y}{>{\centering\arraybackslash}X}

\begin{document}

\title{Requirements Analysis for an \\ Open Research Knowledge Graph}

\author{Arthur Brack\inst{1}\orcidID{0000-0002-1428-5348}
\and Anett Hoppe\inst{1}\orcidID{0000-0002-1452-9509} 
\and \\ Markus Stocker\inst{1}\orcidID{0000-0001-5492-3212} 
\and S\"oren Auer\inst{1,2}\orcidID{0000-0002-0698-2864} 
\and  Ralph~Ewerth\inst{1,2}\orcidID{0000-0003-0918-6297} }

\authorrunning{A. Brack et al.}
\institute{TIB Leibniz Information Centre for Science and Technology, Hannover, Germany \and
L3S Research Center, Leibniz University, Hannover, Germany \\
\email{\{arthur.brack|anett.hoppe|markus.stocker|auer|ralph.ewerth\}@tib.eu}}

\maketitle              % typeset the header of the contribution
\begin{abstract}

Current science communication has a number of drawbacks and bottlenecks which have been subject of discussion lately: Among others, the rising number of published articles makes it nearly impossible to get an overview of the state of the art in a certain field, or reproducibility is hampered by fixed-length, document-based publications which normally cannot cover all details of a research work.
Recently, several initiatives have proposed knowledge graphs (KGs) for organising scientific information as a solution to many of the current issues. The focus of these proposals is, however, usually restricted to very specific use cases. In this paper, we aim to transcend this limited perspective by presenting a comprehensive analysis of requirements for an Open Research Knowledge Graph (ORKG) by (a) collecting daily core tasks of a scientist, (b) establishing their consequential requirements for a KG-based system, (c) identifying overlaps and specificities, and their coverage in current solutions. As a result, we map necessary and desirable requirements for successful KG-based science communication, derive implications and outline possible solutions.

\end{abstract}

\section{Introduction}
Today's scholarly communication is a document-centred process and as such, rather inefficient. Scientists spend considerable time in finding, reading and reproducing research results from PDF files consisting of static text, tables, and figures. The explosion in the number of published articles~\cite{bornmann15growth} aggravates this situation further: It gets harder and harder to stay on top of current research, that is to find relevant works, compare and reproduce them and, later on, to make one's own contribution known for its quality. 

Some of the available infrastructures in the research ecosystem already use \emph{knowledge graphs} (KG)\footnote{Acknowledging that knowledge graph is vaguely defined, we adopt the following definition: A \emph{knowledge graph} (KG) consists of (1) an \emph{ontology} describing a conceptual model, and (2) the corresponding \emph{instance data} following the constraints posed by the ontology.
The construction of a KG involves \emph{ontology design} and \emph{population} with instances.} 
to enhance their services. 
Academic search engines, for instance, such as \textit{Microsoft Academic Knowledge Graph}~\cite{Frber2019TheMA} or \textit{Literature Graph}~\cite{Ammar2018ConstructionOT} employ metadata-based graph structures which link research articles based on citations, shared authors, venues and keywords. 

Recently, initiatives have promoted the usage of KGs in science communication, but on a deeper, semantic level~\cite{auer_soren_2018_1157185,Hars2003StructureOS,pertsas2017scholarly,Manghi_Bardi_Atzori_Baglioni_Manola_Schirrwagen_Principe_2019,Jaradeh2019OpenRK,Oelen2019ComparingRC}. 
They envision the transformation of the dominant document-centred knowledge exchange to knowledge-based information flows by representing and expressing knowledge through semantically rich, interlinked KGs.
Indeed, they argue that a shared structured representation of scientific knowledge has the potential to alleviate some of the science communication's current issues: Relevant research could be easier to find, comparison tables automatically compiled, own insights rapidly placed in the current ecosystem. 
Such a powerful data structure could, more than the current document-based system, also encourage the interconnection of research artefacts such as datasets and source code much more than current approaches (like DOI references etc.); allowing for easier reproducibility and comparison. 
To come closer to the vision of knowledge-based information flows, research articles should be enriched and interconnected through machine-interpretable semantic content. 
Jaradeh et al.'s study~\cite{Jaradeh2019OpenRK} indicates that authors are willing to contribute structured descriptions of their research articles.

However, the work of a researcher is manifold, but current proposals usually focus on a specific use case (e.g. the above-named examples focus on enhancing academic search). In this paper, we provide a detailed analysis of common work tasks in a scientist's daily life and analyse (a) how they could be supported by an ORKG, (b) what requirements result for the design of (b1) the KG and (b2) the surrounding system, (c) how different use cases overlap in their requirements and can benefit from each other. Our analysis is led by the following research questions:

\begin{enumerate}[nosep]
    \item What functionalities should be provided by ORKG interfaces?
    \begin{enumerate}
        \item Which user interfaces are necessary?
        \item Which machine interfaces are necessary? 
    \end{enumerate}
    \item What requirements can be defined for the underlying ontologies? 
    \begin{enumerate}
        \item Which granularity of information representation is needed?
        \item To what degree is domain-specialisation needed?
    \end{enumerate}
    \item What requirements can be defined for the instance data? 
    \begin{enumerate}
        \item Which approaches (human vs. machine) are suitable to populate the KG? 
        \item Which coverage of research artefacts is necessary for the instance data?
        \item Which quality is necessary for the instance data? 
    \end{enumerate}    
\end{enumerate}
We follow the design science research (DSR) methodology~\cite{Hevner2004DesignSI}. In this study, we focus on the first phase of DSR conducting a requirements analysis. The objective is to chart necessary (and desirable) requirements for successful KG-based science communication, and, consequently, provide a map for future research.  

The remainder of the paper is organised as follows. Section 2 summarises related work on research knowledge graphs, scientific ontologies and methods for KG construction. The requirements analysis is presented in Section 3, while Section 4 discusses implications and possible approaches for ORKG construction. Finally, Section 5 concludes the requirements analysis and outlines areas of future work.

\section{Related work}
\label{sec:related_work}
This section provides a brief overview of 
(a) existing research KGs,
(b) ontologies representing scholarly knowledge, and
(c) approaches for KG construction.

\subsection{Research knowledge graphs}

Academic search engines (e.g. Google Scholar, Microsoft Academic, SemanticScholar) exploit graph structures such as the Microsoft Academic Knowledge Graph~\cite{Frber2019TheMA}, SciGraph~\cite{Yaman2019InterlinkingSA}, or the Literature Graph~\cite{Ammar2018ConstructionOT}.
These graphs interlink research articles through metadata, e.g. citations, authors, affiliations, grants, journals, or keywords.

To help reproducing research results, initiatives such as Research Graph~\cite{Amir2017ResearchGB}, Research Objects~\cite{Bechhofer2010WhyLD} and OpenAIRE~\cite{Manghi_Bardi_Atzori_Baglioni_Manola_Schirrwagen_Principe_2019}
interlink research articles with research artefacts such as datasets, source code, software, and presentation videos.
Scholarly Link Exchange (Scholix)~\cite{Burton_2017} aims to create a standardised ecosystem to collect and exchange links between research artefacts and literature.

Some approaches were proposed to interlink articles at a more semantic level:
Paperswithcode.com is a community-driven effort to link machine learning articles with tasks, source code and evaluation results to construct leaderboards.
Ammar et al.~\cite{Ammar2018ConstructionOT} interlink entity mentions in abstracts with DBpedia~\cite{Lehmann2015DBpediaA} and Unified Medical Language System (UMLS)~\cite{Bodenreider2004TheUM}, and Cohan et al.~\cite{Cohan2019StructuralSF} extend the citation graph with semantic citation intents (e.g. cites as background or as used method).

Various scholarly applications benefit from semantic content representation, e.g. academic search engines by exploiting general-purpose KGs~\cite{Xiong2017ExplicitSR},
and graph-based research paper recommendation systems~\cite{Beel2015ResearchpaperRS} by utilising citation graphs and mentioned genes.
However, the coverage of science-specific concepts in general-purpose KGs is rather low \cite{Ammar2018ConstructionOT},
e.g. the task ``geolocation estimation of photos'' from Computer Vision is neither present in Wikipedia nor in CSO (Computer Science Ontology)~\cite{Salatino2019TheCS}.

\subsection{Scientific ontologies}

Various ontologies have been proposed to model metadata such as bibliographic resources and citations~\cite{Peroni2012FaBiOAC}. 
Iniesta and Corcho~\cite{IniestaSurveyOntologies} reviewed ontologies to describe scholarly articles.
In the following, we describe some ontologies that conceptualise the semantic content in research articles. 

Several ontologies focus on rhetorical~\cite{Waard2006TheAF,Groza2006SALTSA,Constantin2016TheDC} (e.g. Background, Methods, Results, Conclusion), 
argumentative~\cite{teufel2009towards,Liakata2010CorporaFT} (e.g. claims, contrastive and comparative statements about other work) or 
activity-based \cite{pertsas2017scholarly} (e.g. sequence of research activities) aspects and elements of research articles.
Others describe scholarly knowledge with interlinked entities such as problem, method, theory, statement~\cite{Hars2003StructureOS,Brodaric2008SKIingWD}, or focus on the main research findings and characteristics of research articles described in surveys with concepts such as problems, approaches, implementations, and evaluations~\cite{Fathalla2017TowardsAK,Vahdati2019SemanticRO}.

There are various domain-specific ontologies, for instance, 
mathematics~\cite{Lange2013OntologiesAL} (e.g. definitions, assertions, proofs) and
machine learning~\cite{Klampanos2018ANNETTOAO,Mesbah2017SemanticAO} (e.g. dataset, metric, model, experiment).
The EXPeriments Ontology (EXPO) is a core ontology for scientific experiments conceptualising experimental design, methodology, and results~\cite{Soldatova2006AnOO}.

Taxonomies for domain-specific research areas support the characterisation and exploration of a research field.
Salatino et al.~\cite{Salatino2019TheCS} provide an overview, e.g. Medical Subject Heading (MeSH), Physics Subject
Headings (PhySH), Computer Science Ontology (CSO).
Gene Ontology~\cite{Consortium2004TheGO} and Chemical Entities of Biological Interest (CheBi)~\cite{Degtyarenko2007ChEBIAD} are KGs for genes and molecular entities.

\subsection{Construction of knowledge graphs}
\label{sec:construction_KG}
\paragraph{Automatic construction from text:}
Petasis et al.~\cite{PetasisKPKZ11} provide a review on \emph{ontology learning}, that is ontology creation from text, while Lubani et al.\cite{LubaniNM19} review \emph{ontology population systems}. Pajura and Singh~\cite{Pujara2018MiningKG} provide an overview of the involved tasks for \emph{KG population}:
(a) \emph{knowledge extraction} to extract a graph from text with \emph{entity extraction} and \emph{relation extraction}, and (b) \emph{graph construction} to clean and complete the extracted graph, as it is usually ambiguous, incomplete and inconsistent.  
\emph{Coreference resolution}~\cite{Luan2018MultiTaskIO} clusters different mentions of the same entity and \emph{entity linking}~\cite{Kolitsas2018EndtoEndNE} maps them to entities in the KG.
For \emph{taxonomy population} Salatino et al.~\cite{Salatino2019TheCS} provide an overview of methods based on rule-based natural language processing (NLP), clustering and statistical methods.
In particular, the Computer Science Ontology (CSO) has been populated automatically from research articles~\cite{Salatino2019TheCS}.

\paragraph{Information extraction from scientific text:}
\label{sec:scientific_information_extraction}
Nasar~et~al.~\cite{Nasar2018InformationEF} provide a survey about scientific information extraction.
Beltagy~et~al.~\cite{Beltagy2019SciBERTPC} present benchmarks for several datasets.

There are datasets which are annotated at \emph{sentence level} for several domains, e.g. biomedical~\cite{Dernoncourt2017PubMed2R,Kim2011AutomaticCO}, computer graphics~\cite{Fisas2015OnTD}, computer science~\cite{Cohan2019PretrainedLM}, chemistry and computational linguistics~\cite{teufel2009towards}. 
They focus either on the rhetorical structure in 
abstracts \cite{Dernoncourt2017PubMed2R,Kim2011AutomaticCO,Cohan2019PretrainedLM} 
or full articles \cite{Fisas2015OnTD,Liakata2010CorporaFT},
or on the argumentative structure of full articles~\cite{teufel2009towards}.
The datasets differentiate between five and twelve concept classes (e.g. Background, Objective, Results).
On abstracts and full articles machine learning approaches achieve an F1 score of 83-92\%~\cite{Cohan2019PretrainedLM} or 51-80\%~\cite{liakata2012automatic,Fisas2015OnTD}, respectively.

More recent corpora, annotated at \emph{phrasal level}, aim at constructing a fine-grained KG from scholarly abstracts with the tasks of concept extraction~\cite{augenstein2017semeval,Luan2018MultiTaskIO,Brack2020DomainindependentEO,handschuh2014acl}, 
relation extraction~\cite{Luan2018MultiTaskIO,gabor2018semeval,augenstein2017semeval},
and coreference-resolution~\cite{Luan2018MultiTaskIO}.
They cover several domains, 
e.g. 
computational linguistics~\cite{gabor2018semeval,handschuh2014acl};
computer science, material sciences, and physics~\cite{augenstein2017semeval};
machine learning~\cite{Luan2018MultiTaskIO};
 or a set of ten scientific, technical and medical domains~\cite{Brack2020DomainindependentEO}.
The datasets differentiate between four to seven concept classes (like Task, Method, Tool) 
and between two to seven relation types (like used-for, part-of, evaluate-for).
Concept extraction, coreference-resolution and relation extraction achieve an F1 score of 45-89\%~\cite{Beltagy2019SciBERTPC,augenstein2017semeval,Brack2020DomainindependentEO}, 
48\%~\cite{Luan2018MultiTaskIO} 
and 28-50\%~\cite{augenstein2017semeval,gabor2018semeval,Luan2018MultiTaskIO}, respectively, 
and the inter-coder agreement is 60-76\%~\cite{augenstein2017semeval,Brack2020DomainindependentEO,Luan2018MultiTaskIO},
68\%~\cite{Luan2018MultiTaskIO} and 
60\%-90\%~\cite{Luan2018MultiTaskIO,gabor2018semeval,augenstein2017semeval,handschuh2014acl}, respectively.
\emph{This indicates, that these tasks are not only difficult for machines but also for humans.}

\paragraph{Manual curation:}
WikiData~\cite{vrandevcic2014wikidata} is one of the most popular KGs with semantically structured, encyclopaedic knowledge curated manually by a community. 
As of March 2020, WikiData comprises 80M entities curated by almost 25.000 active contributors.
The community also maintains a taxonomy of categories and "infoboxes" which define common properties of certain entity types.
Paperswithcode.com is a further community-driven effort to interlink machine learning articles with tasks, source code and evaluation results.
KGs such as Gene Ontology~\cite{Consortium2004TheGO} or Wordnet~\cite{Fellbaum2000WordNetA} are curated by domain experts.
Research article submission portals such as easychair.org enforce the submitter to provide machine-readable metadata.
Librarians and publishers tag new articles with keywords and subjects~\cite{Yaman2019InterlinkingSA}.
Virtual research environments enable the execution of data analysis on interoperable infrastructure and store the data and results in KGs~\cite{Stocker2018TowardsRI}.

\section{Requirements analysis}
As the discussion of related work reveals, 
existing research KGs focus on specific use cases (e.g. improve search engines, help to reproduce research results) and mainly manage metadata and research artefacts about articles.
We envision a KG in which research articles are interlinked through a deep semantic representation of their content to enable further use cases. In the following, we formulate the problem statement and  describe our research method. This motivates our use case analysis in section 3.1, from which we derive requirements for an ORKG. 

\paragraph{Problem statement:}
Scholarly knowledge is very heterogeneous and diverse. 
Therefore, an ontology that conceptualises scholarly knowledge comprehensively does not (and unlikely will) exist.
Besides, due to the complexity of the task, the population of comprehensive ontologies requires domain and ontology experts. 
Current automatic approaches can only populate rather simple ontologies and achieve moderate accuracy (see Section~\ref{sec:construction_KG}).
\emph{On the one hand, we desire an ontology that can comprehensively capture scholarly knowledge and instance data with high quality and coverage. On the other hand, we are faced with a ``knowledge acquisition bottleneck''.}

\paragraph{Research method:}
To illuminate the above problem statement we perform a \emph{requirements analysis}.
We follow the \emph{design science research (DSR)} methodology~\cite{Horvth2007ComparisonOT,Braun2015ProposalFR}. The requirements analysis is a central phase in DSR, as it is
the basis for design decisions and selection of methods to construct effective solutions systematically~\cite{Braun2015ProposalFR}. 
DSR's objective in general is the innovative, rigorous and relevant design of information systems for solving important business problems or the improvement of existing solutions~\cite{Braun2015ProposalFR,Hevner2004DesignSI}.
To elicit requirements, we studied guidelines for systematic literature reviews~\cite{fink2014conducting,Kitchenham07guidelinesfor,Okoli2015AGT} and
interviewed members of the ORKG team at TIB\footnote{https://projects.tib.eu/orkg/project/team/}, who are software engineers and researchers in the field of computer science and environmental sciences.
Based on the requirements, we elaborate possible approaches to construct an ORKG, which were identified through a literature review (see Section~\ref{sec:construction_KG}).
To verify our assumptions on the presented requirements and approaches, ORKG team members reviewed them.

\subsection{Overview of the use cases}

We define functional requirements with use cases~\cite{Booch2015TheUM}. 
A use case describes the inter\-action between a user and the system from the user's perspective to achieve a certain goal. 
As a motivating scenario it also guides the design of a supporting ontology~\cite{Degbelo2017ASO}. 

\begin{figure}[tb]
    \center{\includegraphics[width=\linewidth]
        {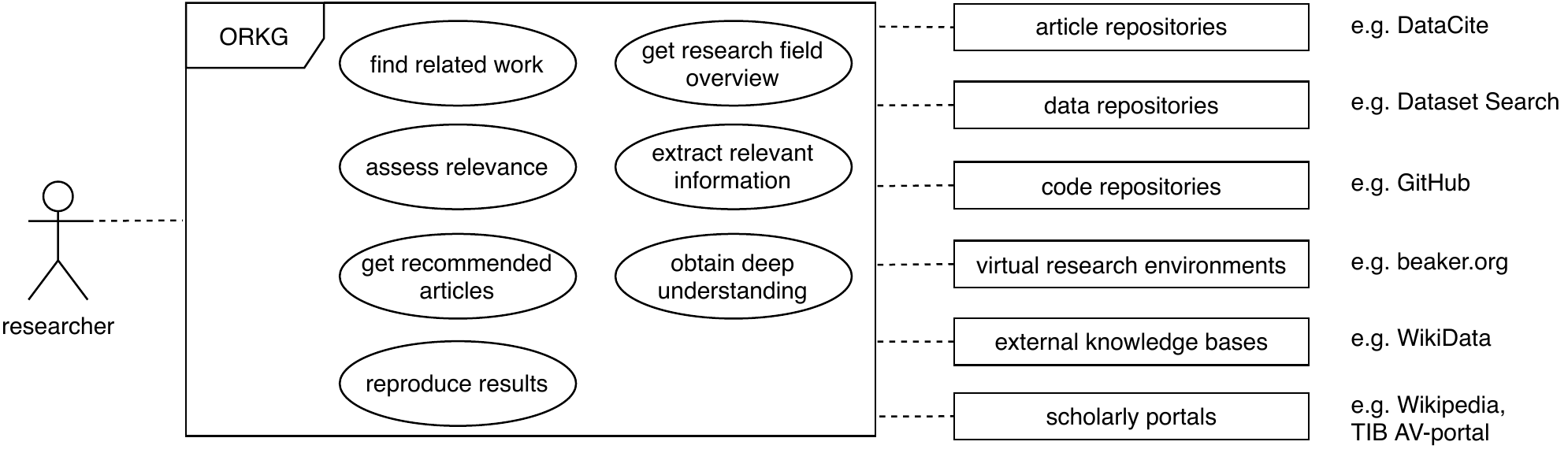}}
    \caption{UML use case diagram for the main use cases between the actor researcher, an Open Research Knowledge Graph (ORKG), and external systems.}
    \label{fig:use_cases}
\end{figure}

There are many use cases (e.g. literature reviews, plagiarism detection, peer reviewer suggestion) and several stakeholders (e.g. researchers, librarians, peer reviewer, practitioners) that may benefit from an ORKG.
In this study, we focus on use cases that support \emph{researchers} (a) conducting literature reviews, (b) obtaining a deep understanding of a research article and (c) reproducing research results. A full discussion of all possible use cases of graph-based knowledge management systems in the research environment is far beyond the scope of this article. With the chosen focus, we hope to cover the most frequent, literature-oriented tasks of scientists. 
Figure~\ref{fig:use_cases} depicts the main identified use cases, which are described briefly in the following. 
Please note that we focus on how \emph{semantic content} can improve these use cases and not further metadata. 

\paragraph{Get research field overview:}
Survey articles provide an overview of a particular research field, e.g. a certain research problem or a family of approaches. 
The results in such surveys are sometimes summarised in structured and comparative tables (an approach usually followed in domains such as computer science, but not as systematically practised in other fields). 
However, once survey articles are published they are no longer updated. 
Moreover, they usually represent only the perspective of the authors, i.e. very few researchers in the field.
To support researchers to obtain an up-to-date overview of a research field, the system should maintain such surveys in a structured way, and allow for dynamics and evolution.
A researcher interested in such an overview should be able to search or to browse the desired research field.
Then, the system should provide related articles and available overviews, e.g. in a table or a leaderboard chart. 
While the user interface shows tabular, leaderboards, or other visual representations the backend should semantically represent information to allow for exploiting overlaps in conceptualisations between research problems or fields.

\paragraph{Find related work:}
Finding relevant research articles is a daily core activity of researchers.
It should be possible to pose queries for related work, which can be fine-grained or broad search intents. 
Systems should preferably support natural language queries as approached by semantic search and question answering engines~\cite{Balog2018EntityOrientedS}. The system has to return a set of relevant articles.

\paragraph{Assess relevance:}
Given a set of relevant articles the researcher has to assess whether the articles match the criteria of interest.
Usually researchers skim through the title and abstract. Sometimes, the introduction and conclusions have to be considered. However, this is usually cumbersome and time-consuming. Presenting the researcher only the most important zones in the article in a structured way can boost this process. This includes, for instance, text passages that describe the problem tackled in the research work, the employed methods or materials, or the yielded results. Also, faceted drill-down methods based on the properties of semantic descriptions of research approaches will empower researchers to quickly filter and zoom into the most relevant literature.

\paragraph{Extract relevant information:}
To tackle a particular research question, the researcher has to extract relevant information from relevant research articles.
Such information is usually compiled in written text or comparison tables in a related work section or survey articles.
For instance, for the question \emph{Which datasets exist for scientific sentence classification?} a researcher who focuses on a new annotation study could be interested in (a) domains covered by the dataset and (b) the inter-coder agreement. 
Another researcher might follow the same question but with a focus on machine learning could be interested in (c) evaluation results and (d) feature types employed. 
The system should support the researcher with tailored information extraction from a set of research articles: (1) the researcher defines a data extraction form as proposed in systematic literature reviews~\cite{Kitchenham07guidelinesfor} (e.g. the above fields (a)-(d)) and (2) the system presents the extracted information for the corresponding data extraction form and articles in a table.

\paragraph{Get recommended articles:}
When the researcher focuses on a particular article, further related articles should be recommended by the system, for instance, articles that address the same research problem or apply similar methods.

\paragraph{Obtain deep understanding:}
The system shall help the researcher to obtain a deep understanding of a research article (e.g. equations, algorithms, diagrams, datasets). 
For this purpose, the system should interlink the article with artefacts such as conference videos, presentations, source code, datasets, etc., and visualise the artefacts appropriately.
Also text passages can be interlinked, e.g. method explanations in Wikipedia, source code snippets implementing algorithms or equations described in the article.

\paragraph{Reproduce results:}
The system should provide the researcher links to all necessary artefacts to reproduce research results, e.g. datasets, source code, virtual research environments, materials describing the study, etc.
Further, the system shall maintain semantic descriptions of domain-specific and standardised evaluation protocols and guidelines.

\subsection{Knowledge graph requirements}
\begin{table}[tb]
\scriptsize
\caption{\small \textbf{Requirements and approaches for the main use cases.} The upper part describes the minimum requirements for the ontology (domain-specialisation and granularity) and the instance data (coverage and quality). The bottom part provides possible approaches for manual, automatic and semi-automatic curation of the KG for the respective use cases. ``X'' indicates that the approach is suitable for the use case while ``(x)'' means that the approach is only appropriate with human supervision. The left part (delimited by the vertical triple line) groups use cases suitable for manual, and the right side for automatic approaches. Vertical double lines group use cases with similar requirements.}        
\label{tab:use_cases_approaches}
\begin{tabularx}{\linewidth}{p{1.2cm}|p{2.6cm}|Y|Y||Y|Y|||Y|Y||Y}
                                 &                           & \textit{Extract relevant info} & \textit{Research field overview} & \textit{Deep understanding} & \textit{Repro\-duce results} & \textit{Find related work} & \textit{Recom\-mend articles} & \textit{Assess relevance} \\\hline \hline
\multirow{2}{*}{\textit{Ontology}}        
                                 & Domain-specialisation                                   & high                        & high                       & med                   & med            & low               & low              & med  \\\cline{2-9}
                                 & Granularity                                             & high                        & high                       & med                    & med           & low               & low              & low  \\\hline
\multirow{2}{*}{\shortstack[l]{\textit{Instance}\\ \textit{data}}}                                                                                                                         
                                 & Coverage                                                & low                         & low                        & low                       & med        & high              & high             & med     \\\cline{2-9}
                                 & Quality                                                 & high                        & high                       & high                     & high        & low               & low              & med       \\\hline\hline\hline
\multirow{4}{*}{\shortstack[l]{\textit{Manual}\\ \textit{curation}}}                                                                                                                          
                                 & Maintain terminologies                                  & -                            & X                           & -                         & -        & X                 & X                & -             \\\cline{2-9}
                                 & Define templates                                        & X                            & X                           & -                         & -        & -                 & -                & -             \\\cline{2-9}
                                 & Fill in templates                                       & X                            & X                           & X                         & X        & -                 & -                & -             \\\cline{2-9}
                                 & Maintain overviews                                      & X                            & X                           & -                         & -        & -                 & -                & -             \\\hline
\multirow{5}{*}{\shortstack[l]{\textit{Automatic}\\ \textit{curation}}}                                                                                                                        
                                 & Entity/relation extraction                              & (x)                            & (x)                         & (x)                         & (x)  & X                 & X                & X                   \\\cline{2-9}
                                 & Entity linking                                          & (x)                          & (x)                         & (x)                       & (x)      & X                 & X                & X             \\\cline{2-9}
                                 & Sentence classification                                 & (x)                          & -                           & (x)                       & -        & -                 & -                & X             \\\cline{2-9}
                                 & Template-based extraction                               & (x)                          & (x)                         & (x)                         & (x)    & -                 & -                & -                 \\\cline{2-9}
                                 & Cross-modal linking                                     & -                            & -                           & (x)                       & (x)      & -                 & -                & -            
\end{tabularx}

\end{table}

The non-functional requirements for the respective use cases are discussed in the light of the following dimensions. 
\begin{enumerate}[nosep] 
    \item \emph{Domain-specialisation of the ontology:} 
    How domain-specific should the concepts be in the ontology?
    Various ontologies (e.g. ~\cite{pertsas2017scholarly,Brack2020DomainindependentEO}) propose domain independent concepts (e.g. Process, Method, Material).
    In contrast, Klampanos et al.~\cite{Klampanos2018ANNETTOAO} present a very domain-specific ontology for artificial neural networks.

    \item \emph{Granularity of the ontology:} 
    Which granularity is required to conceptualise scholarly knowledge?
    For instance, the annotation schemes for scientific corpora (see Section~\ref{sec:construction_KG}) have a rather low granularity, as they do not have more than 10 classes and 10 relation types.
    In contrast, various ontologies (e.g \cite{Hars2003StructureOS,pertsas2017scholarly}) with more than 20-35 classes and over 20-70 relations and properties are fine-grained and have a relatively high granularity.

    \item \emph{Coverage of the instance data:} 
    Given an ontology, to which extent do \emph{all} possible instances in \emph{all} research articles have to be represented in the KG?
    For instance, given an ontology with a class ``Task'', the instance data for that ontology would have a high coverage if all tasks mentioned in all research articles are present.

    \item \emph{Quality of the instance data:} 
    Given an ontology, which quality is necessary for the corresponding instances? 
    In a KG with high quality all present instances must conform to the ontology and reflect the content of the research articles properly, e.g. an article is correctly assigned to the task addressed in the article, the F1 score in the evaluation results is correctly extracted, etc.
\end{enumerate}

Next, we discuss the seven main use cases with regard to the required level of ontology domain-specialisation and granularity, as well as coverage and quality of instance data. 
Table~\ref{tab:use_cases_approaches} summarises the requirements for the use cases along the four dimensions at ordinal scale. 
The use cases are grouped together, when they have (1) similar justifications for the posed requirements, and (2) a high overlap in the ontology concepts and instances.

\paragraph{Extract relevant information \& get research field overview:}
The information to be extracted from relevant research articles for a data extraction form is very heterogeneous and depends highly on the intent of the researcher and the research questions.
Thus, the ontology has to be domain-specific and fine-grained to offer all possible kinds of desirable information.
In addition, the provided information has to be of high quality, e.g. a provided F1 score of an evaluation result must not be wrong.
However, missing information for certain questions in the KG may be tolerable for a researcher.

\paragraph{Obtain deep understanding \& reproduce results:}
The provided information for these use cases has to be of high quality (e.g. accurate links to dataset, source code, videos, articles, research infrastructures). 
The ontology for representing default artefacts can be rather domain-independent (e.g. Scholix~\cite{Burton_2017}).
However, semantic representation of evaluation protocols require domain-dependent ontologies (e.g. EXPO~\cite{Soldatova2006AnOO}).
Missing information is tolerable for these use cases.

\paragraph{Find related work \& get recommended articles:}
When searching for related work, it is essential not to miss relevant articles. 
Previous studies revealed that more than half of search queries in academic search engines refer to scientific entities~\cite{Xiong2017ExplicitSR} and the coverage of scientific entities in KGs is rather low~\cite{Ammar2018ConstructionOT}.
Despite the low coverage, Xiong et al. ~\cite{Xiong2017ExplicitSR} could improve the ranking of search results by exploiting KGs.
Hence, the instance data for the ``find related work'' use case shall have high coverage with fine-grained scientific entities.
However, semantic search engines employ latent representations of KGs and text (e.g. graph and word embeddings)~\cite{Balog2018EntityOrientedS}.
Since a non-perfect ranking of the search results is tolerable for a researcher, lower quality of the instance data is acceptable.
Furthermore, due to latent feature representations, the ontology can be kept rather simple and domain-independent.
For instance, the STM corpus~\cite{Brack2020DomainindependentEO} proposes four domain-independent concepts.
%Graph-based research paper recommendation systems~\cite{Beel2015ResearchpaperRS} also employ latent feature representations and have therefore similar requirements.
Graph- and content-based research paper recommen\-dation systems~\cite{Beel2015ResearchpaperRS} have similar requirements since they also employ latent feature representations, require fine-grained scientific entities, and non-perfect recommendations are tolerable.

\paragraph{Assess relevance:}
To help the researcher to assess the relevance of an article according to her needs, the system should highlight the most essential zones in the article to get a quick overview. 
The coverage and quality of the presented information must not be too low, as otherwise the user acceptance may suffer. 
However, it can be suboptimal, since it is acceptable for a researcher when some of the highlighted information is not essential or when some important information is missing.
The ontology to represent essential information should be rather domain-specific and quite simple (cf. ontologies for scientific sentence classification in Section~\ref{sec:scientific_information_extraction}).

\section{Implications for ORKG construction}

In this section, we discuss the implications for the design and construction of an ORKG and outline possible approaches, which are mapped to the use cases in Table~\ref{tab:use_cases_approaches}.
Based on the discussion in the previous section, we can subdivide the use cases into two groups: 
(1) requiring high quality and high domain-specialisation with only low requirements on the coverage (left side in Table~\ref{tab:use_cases_approaches}), 
and (2) requiring high coverage with rather low requirements on the quality and domain-specialisation (right side in Table~\ref{tab:use_cases_approaches}).
The first group requires manual approaches while the second group could be accomplished with fully automatic approaches. 
However, manually curated data can also support use cases with automatic approaches, and vice versa. Besides, automatic approaches can complement manual approaches by providing suggestions in user interfaces.

\subsection{Manual approaches}

\begin{figure}[t]
    \center{\includegraphics[width=1.0\linewidth]
        {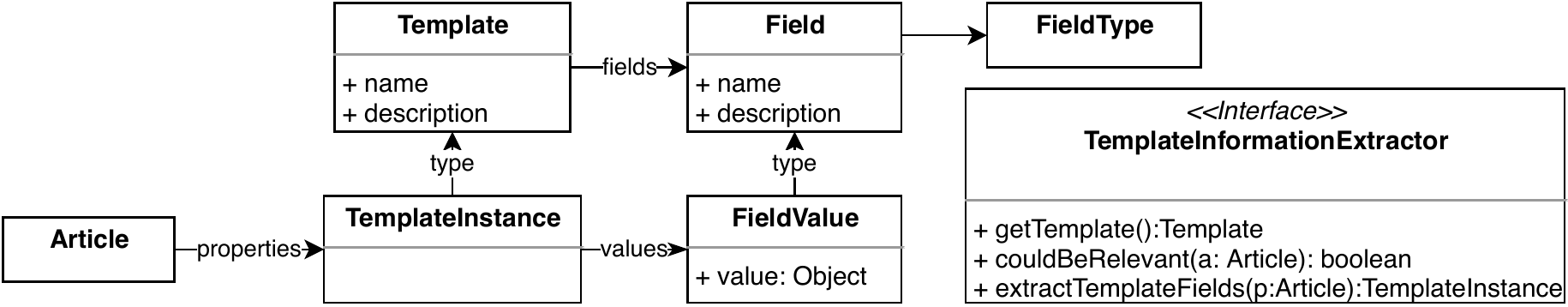}}
    \caption{Conceptual meta-model in UML for templates and interface design for an external template-based information extractor.}
    \label{fig:template_third_party}
\end{figure}

\paragraph{Ontology design:}
The first group of use cases requires rather domain-specific and fine-grained ontologies. 
We suggest to develop novel or reuse ontologies that fit the respective use case and the specific domain (e.g. EXPO~\cite{Soldatova2006AnOO} for experiments).
Moreover, appropriate and simple user interfaces are necessary for efficient and easy population.

However, such ontologies can evolve with the help of the community, as demonstrated by WikiData and Wikipedia with ``infoboxes'' (see Section~\ref{sec:construction_KG}).
Therefore, the system should enable the maintenance of \emph{templates}, which are pre-defined and very specific forms consisting of fields with certain types (see Figure~\ref{fig:template_third_party}).
For instance, to automatically generate leaderboards for machine learning tasks a template would have the fields Task, Model, Dataset and Score, which can then be filled in by a curator for articles providing such kind of results in a user interface generated from the template. 
Such an approach is also called \emph{meta-modelling}~\cite{Booch2015TheUM}, as the meta-model for templates enables the definition of concrete templates, which are then instantiated for articles.

\paragraph{Knowledge graph population:}
Several user interfaces are required to enable manual population:
(1) populate semantic content for a research article by (1a) choosing relevant templates or ontologies and (1b) fill in the values;
(2) terminology management (e.g. domain-specific research fields);
(3) maintain research field overviews by (3a) assigning relevant research articles to the research field, (3b) define corresponding templates and (3c) fill in the templates for the relevant research articles.

Further, the system should also provide \emph{APIs} to enable population by third-party applications, e.g. 
(i) submission portals such as easychair.org during submission of an article; 
(ii) authoring tools such as overleaf.com during writing;
(iii) virtual research environments~\cite{Stocker2018TowardsRI} to store evaluation results and links to datasets and source code during experimenting and data analysis.

To \emph{encourage crowd-sourced content}, we see the following options:
(a) \emph{top-down enforcement} via submission portals and publishers.
(b) \emph{Incentive models}: Researchers want their articles to be cited. Semantic content helps other researchers to find, explore and understand an article. 
(c) Provide \emph{public acknowledgements} for curators.

\subsection{(Semi-)automatic approaches}
The second group of use cases require a high coverage while a rather low quality and domain-specialisation are acceptable.
For these use cases, rather simple and domain-independent ontologies should be developed or reused.

Various approaches can be employed to populate an ORKG (semi-)automatically.
Methods for \emph{entity and relation extraction} (see Section~\ref{sec:construction_KG}) can help to populate fine-grained KGs with high coverage and \emph{entity linking} approaches can link mentions in text with entities.
For cross-modal linking, Singh et al.~\cite{Singh2016OCRAR} propose an approach to detect URLs to datasets in research articles automatically, while the Scientific Software Explorer~\cite{Hoppe2018AnAT} interlinks text passages in research articles with code fragments.
To extract relevant information at sentence level, approaches for \emph{sentence classification} in scientific text can be employed (see Section~\ref{sec:construction_KG}). 
To support the curator fill in templates semi-automatically, \emph{template-based extraction} can (1) suggest relevant templates for a research article and (2) pre-fill fields of templates with appropriate values. 
For pre-filling, approaches such as for natural language inference used in leaderboard construction~\cite{Hou2019IdentificationOT} or end-to-end question answering~\cite{Rajpurkar2016SQuAD10,Devlin2018BERTPO} can be employed.

Further, the system shall enable to plugin \emph{external information extractors}, developed for certain scientific domains to extract specific types of information. For instance, as depicted in Figure~\ref{fig:template_third_party}, an external template information extractor has to implement an interface with three methods. This enables the system (1) to filter relevant template extractors for an article and (2) extract field values from an article.

\section{Conclusions}

In this paper, we have presented a requirements analysis for an Open Research Knowledge Graph (ORKG). 
An ORKG shall represent the content of research articles in a semantic way to enhance or enable a wide range of use cases.
We identified literature-related core tasks of a researcher that can be supported by an ORKG and formulated them as use cases. 
For each use case, we discussed specificities and requirements for the underlying ontology and the instance data.
In particular, we identified two groups of use cases: 
(1) the first group requires high-quality instance data and rather fine-grained, domain-specific ontologies, but with moderate coverage;
(2) the second group requires a high coverage, but the ontologies can be kept rather simple and domain-independent, and a moderate quality of the instance data is sufficient.
Based on the requirements, we have described possible manual and semi-automatic approaches (necessary for the first group), and automatic approaches (appropriate for the second group) for KG construction. 
In particular, we propose a framework with lightweight ontologies that can evolve by community curation.
Further, we have described the interdependence with external systems, user interfaces, and APIs for third-party applications to populate an ORKG.

The results of our work aim to provide a holistic view of the requirements for an ORKG and be a guideline for further research. 
The suggested approaches have to be refined, implemented and evaluated in an iterative and incremental process (see www.orkg.org for the current progress). 
Additionally, our paper can serve as a foundation for a discussion on ORKG requirements with other researchers and practitioners. 

\bibliographystyle{splncs04}
\bibliography{references}

\end{document}